\documentstyle[preprint,aps]{revtex} 

\begin{document}

\draft 
\preprint{MSUPHY99.05} 
\title{Semiclassical charged black holes with a	quantized massive
scalar field}

\author{Brett E.\ Taylor\cite{BTA,BTE}, William A.\ Hiscock\cite{BH}}

\address{Department of Physics, Montana State University, Bozeman,
Montana 59717 }

\author{Paul R. Anderson \cite{PRA}}

\address{Department of Physics, Wake Forest University,
Winston-Salem, North Carolina 27109}

\date{November 29, 1999} 
\maketitle 


\begin{abstract} 
Semiclassical perturbations to the Reissner-Nordstr\"{o}m metric
caused by the presence of a quantized massive scalar field with
arbitrary curvature coupling are found to first order in $\epsilon =
\hbar/M^2$.  The DeWitt-Schwinger approximation is used to determine
the vacuum stress-energy tensor of the massive scalar field. When the
semiclassical perturbation are taken into account, we find
extreme black holes will have a charge-to-mass ratio that exceeds
unity, as measured at infinity.  The effects of the perturbations on
the black hole temperature (surface gravity) are studied in detail,
with particular emphasis on near extreme ``bare'' states that might
become precisely zero temperature ``dressed'' semiclassical black hole
states.  We find that for minimally or conformally coupled scalar
fields there are {\it no} zero temperature solutions among the
perturbed black holes.

\end{abstract} 

\section{Introduction}
\label{sec:intro}
 
The backreaction of quantized fields on the spacetime geometry of a
black hole can have very significant and important implications.
For example backreaction from the particle production that occurs
in the Hawking effect~\cite{Hawking} causes the black hole to gradually
become smaller in size.  As it does so its temperature becomes higher
and its entropy lower.  If a black hole is placed in thermal equilibrium
with radiation in a cavity the backreaction of quantized fields
can again alter its temperature and entropy~\cite{York,HKY,AHWY}.

Due to the difficulty in computing the stress-energy tensor for
quantized fields in black hole spacetimes various approximations have
been used for all backreaction calculations that have been done so
far.  One extremely useful approximation is to calculate (either
exactly, or within some approximation scheme) $\langle T^{\mu}\!_{\nu}
\rangle$ in a classical black hole geometry, and then compute
semiclassical corrections to the metric as linear perturbations.  This
works particularly well for static solutions such as occur for a zero
temperature black hole or a black hole in thermal equilibrium with
radiation in a cavity.  The fact that the solutions are static makes
the problem much more tractable.  An advantage of this approach is
that it gives direct information about how semiclassical effects alter
the geometry from that of the corresonding classical solution to
Einstein's equations.  Even though such models have so far only been
applied to the static case, they are relevant to the issue of the end
point of black hole evaporation, because the perturbations give
information about how quantum effects alter the temperature of a black
hole.  For example, if uncharged zero temperature solutions were found
which possessed nonzero mass, these could be potential end-point
``remnants'' of black hole evaporation.

To date the linearized semiclassical backreaction equations have only
been solved in the case of an initially Schwarzschild black hole in
thermal equilibrium with massless radiation in a cavity.  York
considered the perturbation due to a conformally coupled quantized
scalar field \cite{York}.  Hochberg, Kephart, and York extended this
work to include the effects of massless quantized spinor and vector
fields \cite {HKY}.  Anderson Hiscock, Whitesell, and York \cite{AHWY}
studied the perturbations due to the vacuum stress-energy of a
quantized massless scalar field with arbitrary curvature coupling.
In all these cases, the stress-energy tensor of the quantized field
was treated using analytic approximations developed by Page, Brown,
and Ottewill \cite{P,BO,PBO} and Anderson, Hiscock, and Samuel 
\cite{AHS}.

In this paper we investigate charged and uncharged black holes
which interact with
an uncharged quantized massive scalar field of arbitrary curvature
coupling.  The general ``bare'' spacetime is described by the
Reissner-Nordstr\"{o}m metric:
\begin{equation}
	ds^2 = -\left(1 - \frac{2M}{r} + \frac{Q^2}{r^2}\right)dt^2 + 
		\left(1- \frac{2M}{r} + \frac{Q^2}{r^2}\right)^{-1} dr^2 + 
		r^2 d\Omega ^2 \; ,
\label{rn_metric}
\end{equation}
where $Q$ is the charge on the black hole and $M$ is its mass.  We
treat the vacuum stress-energy of the quantized massive scalar field
as a perturbation on the ``bare'' Reissner-Nordstr\"{o}m spacetime,
solving the semiclassical Einstein equations to find the first-order
in $\hbar$ semiclassical corrections to the Reissner-Nordstr\"{o}m
metric.  We consider the situation analogous to that of
Ref.\cite{AHWY}, where the black hole is in thermal equilibrium with
the quantized field, imposing microcanonical boundary conditions on a
spherical boundary surface surrounding the black hole.  The vacuum
stress-energy is analytically approximated using the DeWitt-Schwinger
approximation; previous exact, numerical work by Anderson, Hiscock,
and Samuel \cite{AHS} (hereafter, AHS) has demonstrated that the
DeWitt-Schwinger approximation to the vacuum stress-energy is quite
good (one percent or better) in the Reissner-Nordstr\"{o}m spacetime
when $mM \geq 2$.

The perturbation caused by the quantized field will change the
temperature of the black hole; we examine in detail the sign and size
of this effect.  We are particularly interested in situations where
the perturbed black hole has precisely zero temperature.  Within the
context of a perturbative approach, this means the unperturbed
Reissner-Nordstr\"{o}m spacetime must be nearly extreme.  The
``bare'' Reissner-Nordstr\"{o}m spacetime that results in a
``dressed'' zero temperature black hole could be either a nearly
extreme Reissner-Nordstr\"{o}m black hole, or possibly a
Reissner-Nordstr\"{o}m naked singularity, with $|Q|$ slightly greater
than $M$.  We are able to handle the case of ``bare'' naked
singularities because the DeWitt-Schwinger approximation for $\langle
T^{\mu}\!_{\nu} \rangle$ is purely local.

Section \ref{sec:background} describes the approximate vacuum stress-energy
tensor and the semiclassical linearized Einstein equations. The metric 
perturbations are also derived and displayed in this section.  In Section
\ref{sec:results} the temperature perturbations are determined, and we
search for zero temperature solutions. We find that there are no 
zero temperature solutions for plausible values of the scalar field's
curvature coupling; specifically, there are none for minimally or conformally
coupled scalar fields.  Section \ref{sec:summary} summarizes our conclusions.
Throughout this paper we use units such that $\hbar = G = c = k_B = 1$.
The sign conventions are those of Misner, Thorne, and Wheeler \cite{MTW}.


\section{Semiclassical Perturbation Method}
\label{sec:background}

In semiclassical gravity, one quantizes the matter fields but not the
spacetime geometry. This modifies the right hand side of the Einstein
field equations, replacing the classical stress-energy tensor with the
expectation value of the quantum stress-energy operator.
The semiclassical Einstein equations then take the form:
\begin{equation}
        G^{\mu}\!_{\nu} = 8 \pi \langle  T^{\mu}\!_{\nu} \rangle \; .
\label{QEFE}
\end{equation}
In examining the semiclassical perturbations of the
Reissner-Nordstr\"{o}m metric caused by the vacuum energy of a
quantized scalar field, we will continue to consider the
electromagnetic field to be classical; for the Reissner-Nordstr\"{o}m
geometry, then, the semiclassical equations will contain both
classical and quantum stress-energy contributions,
\begin{equation}
        G^{\mu}\!_{\nu} = 8 \pi \left[ \langle  T^{\mu}\!_{\nu} \rangle 
        + T^{\mu}\!_{\nu} \right] \; .
\label{QEFE2}
\end{equation}
The classical stress-energy term on the right-hand side of Eq.\
(\ref{QEFE2}) represents the electromagnetic field's stress-energy in
the Reissner-Nordstr\"{o}m geometry; it will vanish in the
Schwarzschild limit as $Q \rightarrow 0$.

The exact calculation of the expectation value for the stress-energy
of a quantized field in a curved spacetime is a non-trivial exercise.
Anderson, Hiscock, and Samuel \cite{AHS} have developed a method for
numerically calculating the vacuum stress-energy tensor of both
massive and massless quantized scalar fields with arbitrary curvature
coupling in a general static, spherically symmetric spacetime.  As
part of this method, they also developed an analytic approximation to
$\langle T^{\mu}\!_{\nu} \rangle$ for massive scalar fields based on
the DeWitt-Schwinger expansion in inverse powers of the field mass.
This approximation is state-independent and entirely local, depending
at each point only on the values of the curvature and its derivatives.
One expects, a priori, the approximation to become increasingly
accurate as the ratio of the Compton wavelength of the field to the
local radius of curvature approaches zero, i.e., in the limit
$ Mm \gg 1$.

They then applied these techniques to the Reissner-Nordstr\"{o}m
spacetime, obtaining exact numerical values for the vacuum
stress-energy for both massive and massless fields.  Comparing the
exact values of $\langle T^{\mu}\!_{\nu} \rangle$ to those provided by
the DeWitt-Schwinger approximation, they showed that in the
Reissner-Nordstr\"{o}m spacetime, the approximate values are quite
good (within a few percent of the exact values) near the horizon if
the field mass is chosen to satisfy $ mM \geq 2$.  One might expect
the DeWitt-Schwinger approximation to also fail to be adequate for
black holes with nonzero temperature, at large values of $r$ where
temperature-dependent (hence, state-dependent) terms associated with
the gas of produced particles dominate $\langle T^{\mu}\!_{\nu}
\rangle$.  However, for a massive field, there will be essentially
no particles created by the hole if the temperature is substantially
less than the mass of the field, $ T \ll m$.  For any
Reissner-Nordstr\"{o}m black hole, the temperature satisfies $ T <
(4\pi M)^{-1}$, so the temperature will be substantially less than the
field mass so long as $ (4\pi)^{-1} \ll Mm$.  This condition is
adequately satisfied when the previously mentioned criterion for the
validity of the DeWitt-Schwinger approximation, $ mM \gg 2$ holds.
Hence, for any Reissner-Nordstr\"{o}m black hole and massive scalar
field combination satisfying $ mM \gg 2$, we will assume the 
DeWitt-Schwinger approximation is valid throughout the region exterior 
to the horizon.
\footnote{Strictly speaking the expectation value of the
stress-energy tensor for a massive field in a thermal state must
approach a nonzero constant in the limit $r \rightarrow \infty$.
However in the case when $mM \gg 1$ there will be very few
"realizations" of the quantum field theory in which any particles are
present.  Another way to think about this is that in a nonstatic state
it would take a very long time on the average before a particle is
produced.  Thus to a good approximation it should be adequate to use
the DeWitt-Schwinger approximation out to arbitrarily large values of
$r$.  Of course for the zero temperature case one expects the
DeWitt-Schwinger approximation to be valid throughout the region
exterior to the event horizon so long as $mM \gg 1$ is satisfied.}

Using the results of AHS for the case of a quantized massive scalar
field in the Reissner-Nordstr\"{o}m spacetime the following values for
the expectation value of the stress-energy tensor are obtained:
\begin{eqnarray}
        \langle T^{t}\!_{t} \rangle & = & 
	\frac{\epsilon}{\pi ^2 m^2}\left[\frac{1237 M^5}{5040 r^9} - 
	\frac{25 M^4}{224 r^8} - \frac{1369 M^5 r_{+}}{1008 r^{10}} + 
	\frac{41 M^4 r_{+}}{105 r^9} + \frac{3 M^3 r_{+}}{56 r^8}  
	\right. \nonumber \\ \nonumber \\
	& &{} \left.
	+\frac{613 M^5 r_{+}^2}{210 r^{11}} - \frac{73 M^4 r_{+}^2}{3360 r^{10}}
	- \frac{41 M^3 r_{+}^2}{210 r^9} - \frac{3 M^2 r_{+}^2}{112 r^8} -
	\frac{2327 M^5 r_{+}^3}{1260 r^{12}} - \frac{613 M^4 r_{+}^3}
	{210 r^{11}}  
	\right. \nonumber \\ \nonumber \\
	& &{} \left.
	+\frac{883 M^3 r_{+}^3}{1260 r^{10}} + \frac{2327 M^4 r_{+}^4}
	{840 r^{12}} + \frac{613M^3 r_{+}^4}{840 r^{11}} - 
	\frac{883 M^2 r_{+}^4}{5040 r^{10}} - \frac{2327 M^3 r_{+}^5}
	{1680 r^{12}} 
	\right. \nonumber \\ \nonumber \\
	& &{} \left.
	+ \frac{2327 M^2 r_{+}^6}{10080 r^{12}}
	+\xi \left(\frac{-11 M^5}{10r^9} + \frac{M^4}{2 r^8} + 
	\frac{217 M^5 r_{+}}{30 r^{10}} - \frac{14M^4 r_{+}}{5 r^9} - 
	\frac{226 M^5 r_{+}^2}{15 r^{11}} 
	\right. \right. \nonumber \\ \nonumber \\
	& &{} \left. \left.
	+ \frac{77M^4 r_{+}^2}{180 r^{10}} + \frac{7 M^3 r_{+}^2}{5 r^9}
	+ \frac{91M^5 r_{+}^3}{10r^{12}} + \frac{226M^4 r_{+}^3}{15r^{11}} 
	- \frac{182M^3 r_{+}^3}{45 r^{10}} - \frac{273 M^4 r_{+}^4}{20r^{12}}
	\right. \right. \nonumber \\ \nonumber \\  
	& &{} \left. \left. 
	- \frac{113M^3 r_{+}^4}{30 r^{11}} + \frac{91M^2 r_{+}^4}{90r^{10}}
	+\frac{273 M^3 r_{+}^5}{40 r^{12}} - \frac{91M^2 r_{+}^6}{80 r^{12}} 
	\right) \right] \; ,
\label{Ttt_ds}
\end{eqnarray}
\begin{eqnarray}
        \langle T^{r}\!_{r} \rangle & = & 
	\frac{\epsilon}{\pi ^2 m^2}\left[ \frac{-47M^5}{720r^9} + 
	\frac{7M^4}{160r^4} + \frac{2081M^5 r_{+}}{5040r^{10}} - 
	\frac{16M^4 r_{+}}{63 r^9} + \frac{3M^3 r_{+}}{280 r^8}
	\right. \nonumber \\ \nonumber \\
	& &{} \left.
	- \frac{13M^5 r_{+}^2}{18r^{11}}
	+\frac{983M^4 r_{+}^2}{10080r^{10}} + \frac{8 M^3 r_{+}^2}{63 r^9} -
	\frac{3M^2 r_{+}^2}{560r^8} + \frac{421M^5 r_{+}^3}{1260r^{12}} + 
	\frac{13M^4 r_{+}^3}{18r^{11}} 
	\right. \nonumber \\ \nonumber \\
	& &{} \left.
	- \frac{383 M^3 r_{+}^3}{1260r^{10}}
	- \frac{421M^4 r_{+}^4}{840r^{12}}
	-\frac{13M^3 r_{+}^4}{72r^{11}} + \frac{383M^2 r_{+}^4}{5040r^{10}} +
	\frac{421M^3 r_{+}^5}{1680 r^{12}} 
	\right. \nonumber \\ \nonumber \\
	& &{} \left.
	- \frac{421M^2 r_{+}^6}{10080r^{12}}
	+\xi \left(\frac{3M^5}{10r^9} - \frac{M^4}{5r^8} - 
	\frac{49M^5 r_{+}}{30r^{10}} + \frac{14M^4 r_{+}}{15r^9} + 
	\frac{14M^5 r_{+}^2}{5 r^{11}} 
	\right. \right. \nonumber \\ \nonumber \\
	& &{} \left. \left.
	- \frac{61M^4 r_{+}^2}{180 r^{10}}
	- \frac{7M^3 r_{+}^2}{15r^9}
	- \frac{13M^5 r_{+}^3}{10r^{12}} - \frac{14M^4 r_{+}^3}{5r^{11}} + 
	\frac{52M^3 r_{+}^3}{45r^{10}} + \frac{39M^4 r_{+}^4}{20r^{12}}  
	\right. \right. \nonumber \\ \nonumber \\
	& &{} \left. \left.
	+ \frac{7M^3 r_{+}^4}{10r^{11}}
	- \frac{13M^2 r_{+}^4}{45r^{10}}
	- \frac{39M^3 r_{+}^5}{40r^{12}} + \frac{13M^2 r_{+}^6}{80r^{12}}
	\right) \right] \; ,
\label{Trr_ds}
\end{eqnarray}
where $r_+ = M + \sqrt{M^2-Q^2}$ is the radius of the unperturbed
event horizon, $\epsilon = 1/M^2$ is our expansion parameter for the
perturbation (in conventional units, $\epsilon = M^{2}_{\rm
Planck}/M^2$) and $\xi$ is the curvature coupling for the field.  We
do not display the value of $\langle T^{\theta}\!_{\theta}
\rangle$, as it not needed here.  Knowledge of the two components
shown above is sufficient to completely solve the perturbed
semiclassical Einstein equations.

The semiclassical Einstein equations may be more easily solved if a
coordinate transformation is made to ingoing Eddington-Finklestein
coordinates.  The transformation is described by
\begin{eqnarray}
        \frac{\partial t}{\partial v} & = &1 \; , \\
	\frac{\partial t}{\partial \tilde{r}} &=& 
		- \left(1 -\frac{2M}{\tilde{r}} + 
		\frac{Q^2}{\tilde{r}^2} \right)^{-1} \; ,\\
	\frac{\partial r}{\partial v} &=& 0 \; ,\\
	\frac{\partial r}{\partial \tilde{r}} &=& 1 \; .
\label{ef_transform}
\end{eqnarray}
The Reissner-Nordstr\"{o}m metric takes the following form in
ingoing Eddington-Finklestein coordinates
\begin{equation}
        ds^2 = -\left(1 -\frac{2M}{\tilde{r}} + \frac{Q^2}{\tilde{r}^2} \right)
		dv^2 + 2dv d\tilde{r} + \tilde{r}^2 d\Omega ^2 \; .
\label{rn_efcoords}
\end{equation}
The components of the expectation value of the stress-energy tensor in 
these coordinates are:
\begin{eqnarray}
        \langle T^v\!_v \rangle &=& \langle T^t\!_t \rangle \; ,
\label{Tvv_ef} \\
	\langle T^{\tilde{r}}\!_{\tilde{r}} \rangle &=& \langle T^r\!_r 
		\rangle \; , 
\label{Trr_ef} \\
        \langle T^v\!_{\tilde{r}}\rangle &=& \left(1 - \frac{2M}{\tilde{r}} +
		\frac{Q^2}{\tilde{r}^2}\right)^{-1}\left[\langle T^r\!_r \rangle
		- \langle T^t\!_t\rangle \right] \; .
\label{Tvr_ef}
\end{eqnarray}

Setting $\tilde{r} = r$, one can write the metric for
a general static spherically symmetric spacetime as
\begin{equation}
	ds^2 = -{\rm e} ^{2 \psi (r)}\left(1 - \frac{2m(r)}{r} +
		\frac{Q^2}{r^2} \right) dv^2 + 2 {\rm e} ^{\psi (r)}dv dr
		+ r^2 d\Omega ^2 \; .
\label{pert_metric_efcoords}
\end{equation}
The perturbations to the Reissner-Nordstr\"{o}m metric can be
introduced by an expansion of the ${\rm e}^\psi$ and $m(r)$ metric
functions to first order in the parameter $\epsilon$:
\begin{mathletters}
\begin{eqnarray}
        {\rm e} ^{\psi (r)} &=& 1 + \epsilon \rho (r)  \; ,
\label{psi}
\end{eqnarray}
\begin{eqnarray}
	m(r) &=& M[1 + \epsilon \mu (r)] \; .
\label{mass_func}
\end{eqnarray}
\end{mathletters}
The components of the Einstein tensor can now be calculated in the metric given 
by Eq.\ (\ref{pert_metric_efcoords}) with the expansions given in Eqs.\ 
(\ref{psi},\ref{mass_func}).  These can be substituted into Eq.\ (\ref{QEFE}) 
along with the classical background stress-energy and the approximate
stress-energy of the quantized field from Eqs.\ (\ref{Tvv_ef} - \ref{Tvr_ef}).
This yields two first order differential equations for $\mu(r)$ and 
$\rho(r)$:
\begin{eqnarray}
        \frac{d \mu}{dr} &=& - \frac{4\pi r^2}{M \epsilon}
		\langle T^t\!_t \rangle  \; , 
\label{mu_eqn} \\ \nonumber \\
	\frac{d \rho}{d r} &=& \frac{4 \pi r}{\epsilon}
		\left(1 - \frac{2M}{r} + \frac{Q^2}{r^2} \right)^{-1}
		\left[\langle T^r\!_r \rangle - \langle T^t\!_t \rangle \right]
		\; .
\label{rho_eqn}
\end{eqnarray}
Here $\langle T^{t}\!_{t}\rangle$ and $\langle T^{r}\!_{r}\rangle$ are
given by Eq.\ (\ref{Ttt_ds}) and Eq.\ (\ref{Trr_ds}).  The $\epsilon$
factors in the denominator of the leading terms in both Eq.\
(\ref{mu_eqn}) and Eq.\ (\ref{rho_eqn}) are exactly canceled by the
overall factor of $\epsilon$ in the expressions for $\langle T^t\!_t
\rangle$ and $\langle T^r\!_r \rangle$.  These differential equations
can be integrated to find the general solutions for $\mu$ and $\rho$.
They are
\begin{eqnarray}
\mu &=& C_1 + \frac{1}{\pi m^2}\left[\frac{1237 M^4}{7560 r^6} - 
        \frac{5M^3}{56r^5} - \frac{4169 M^4}{158760 r_{+}^6} + 
	\frac{461 M^3}{6615 r_{+}^5} - \frac{6607 M^2}{105840 r_{+}^4} + 
	\frac{3007 M}{158760 r_{+}^3}
\right. \nonumber \\ \nonumber \\
& & {} \left.
        - \frac{1369 M^4 r_{+}}{1764 r^7} + \frac{82M^3 r_{+}}{315 r^6} +
	\frac{3 M^2 r_{+}}{70 r^5} + \frac{613 M^4 r_{+}^2}{420 r^8} -
	\frac{73 M^3 r_{+}^2}{5880 r^7} - \frac{41 M^2 r_{+}^2}{315 r^6} 
\right. \nonumber \\ \nonumber \\
& & {} \left.
        - \frac{3 M r_{+}^2}{140 r^5}
	- \frac{2327 M^4 r_{+}^3}{2835 r^9} - \frac{613 M^3 r_{+}^3}{420r^8} +
	\frac{883 M^2 r_{+}^3}{2205 r^7} + \frac{2327M^3 r_{+}^4}{1890 r^9} +
	\frac{613 M^2 r_{+}^4}{1680 r^8}
\right. \nonumber \\ \nonumber \\
& & {} \left.
        - \frac{883 M r_{+}^4}{8820 r^7}
	- \frac{2327M^2 r_{+}^5}{3780 r^9} + \frac{2327M r_{+}^6}{22680 r^9} +
	\xi \left(-\frac{11M^4}{15 r^6} + \frac{2M^3}{5r^5} + 
	\frac{4 M^4}{45 r_{+}^6} - \frac{11M^3}{45r_{+}^5}  
\right. \right. \nonumber \\ \nonumber \\
& & {} \left. \left.
        + \frac{41 M^2}{180 r_{+}^4} - \frac{13 M}{180 r_{+}^3}
	+ \frac{62 M^4 r_{+}}{15r^7} - \frac{28 M^3 r_{+}}{15r^6} -
	\frac{113 M^4 r_{+}^2}{15r^8} + \frac{11 M^3 r_{+}^2}{45 r^7} + 
	\frac{14 M^2 r_{+}^2}{15 r^6} 
\right. \right. \nonumber \\ \nonumber \\
& & {} \left. \left.
        + \frac{182 M^4 r_{+}^3}{45r^9} + \frac{113 M^3 r_{+}^3}{15 r^8}
	- \frac{104M^2 r_{+}^3}{45 r^7} - \frac{91 M^3 r_{+}^4}{15 r^9} -
	\frac{113M^2 r_{+}^4}{60r^8} + \frac{26 M r_{+}^4}{45 r^7}  
\right. \right. \nonumber \\ \nonumber \\
& & {} \left. \left.
        + \frac{91 M^2 r_{+}^5}{30 r^9} - \frac{91 M r_{+}^6}{180r^9}
	\right) \right] \; ,
\label{mu}
\end{eqnarray}
and
\begin{eqnarray}
\rho &=& C_2 + \frac{1}{\pi m^2} \left[ -\frac{29M^4}{280r^6} + 
        \frac{817M^4}{3528r_{+}^6} - \frac{3221 M^3}{8820r_{+}^5} + 
	\frac{253 M^2}{1680 r_{+}^4} + \frac{184 M^4 r_{+}}{441 r^7} + 
	\frac{M^3 r_{+}}{35r^6}
\right. \nonumber \\ \nonumber \\
& & {} \left.
        -\frac{229M^4 r_{+}^2}{420r^8} - \frac{92M^3 r_{+}^2}{441 r^7} -
	\frac{M^2 r_{+}^2}{70 r^6} + \frac{229M^3 r_{+}^3}{420r^8} -
	\frac{229M^2 r_{+}^4}{1680r^8}
\right. \nonumber \\ \nonumber \\
& & {} \left.
        + \xi \left( \frac{7M^4}{15r^6} - \frac{14M^4}{15r_{+}^6} + 
	\frac{23M^3}{15 r_{+}^5} - \frac{13M^2}{20r_{+}^4} - 
	\frac{32M^4 r_{+}}{15r^7} + \frac{13 M^4 r_{+}^2}{5 r^8} + 
	\frac{16M^3 r_{+}^2}{15 r^7} 
\right. \right. \nonumber \\ \nonumber \\
& & {} \left. \left.
        - \frac{13M^3 r_{+}^3}{5r^8} + \frac{13M^2 r_{+}^4}{20r^8}
	\right) \right] \; .
\label{rho}
\end{eqnarray}
The integration constants for both $\mu$ and $\rho$ have been chosen so
that $\mu(r_+) = C_1$ and $\rho(r_+) = C_2$.  The perturbed spacetime is now 
defined to first order in $\epsilon$ to within the two integration constants 
$C_1$ and $C_2$.

The horizon radius is no longer located at $r_+$ due to the perturbation from 
the presence of the quantized field.  Its radius is now defined implicitly
as the solution to the equation 
\begin{equation}
	r_h = m(r_h) + \sqrt{m(r_h)^2 - Q^2} \; .
\label{r_h_initial}
\end{equation}
We can utilize the horizon location to define the perturbed mass of the
black hole,
\begin{equation}
        M_{BH} = m(r_h) = M\left[1 + \epsilon \mu(r_+)\right] \; ,
\label{M_bh_two}
\end{equation}
to first order in $\epsilon$; $r_h$ has been changed to $r_+$ in the
final expression on the right, as the difference would be of order
$\epsilon^2$.  This physical, or dressed, mass, 
$M_{BH}$, is a function of the bare, and unmeasurable mass, $M$, 
plus a small perturbation:
\begin{equation}
        M_{BH} = M + \epsilon M C_{1} \; .
\label{M_bh_one}
\end{equation}
The horizon radius is then expressed in terms of the dressed mass of the
black hole
\begin{equation}
        r_{h} = M_{BH} + \sqrt{M_{BH}^2 - Q^2} \; .
\label{r_h_two}
\end{equation}
The bare mass, $M$, and the mass perturbation, $\delta M$, cannot
be measured independently; only the dressed mass $M_{BH}$ has
physical meaning. We will hereafter only refer to the dressed mass,
$M_{BH}$, defined implicitly in Eq.\ (\ref{r_h_two}). The arbitrary 
but physically unmeasureable integration constant $C_{1}$ is 
then absorbed into the definition of $M$, as in Ref.\cite{AHWY}.
The perturbed metric's mass function now takes the form
\begin{equation}
m(r) = M_{BH} \left[ 1 + \epsilon \tilde{\mu} (r) \right] \; ,
\label {newm}
\end{equation}
where
\begin{equation}
\tilde{\mu} (r) = \mu (r) |_{C_{1}=0} \; .
\label{tildemu}
\end{equation}
The metric can now be rewritten in terms of the dressed mass $M_{BH}$ and 
$\tilde{\mu} (r)$.  Because these quantities are those that can 
be physically measured, the $BH$ subscript and the tilde will now be dropped, 
writing only $M$ and $\mu$ respectively. In addition, we will
denote $r_h$ by $r_+$ henceforth, since they have the same
definition once the mass has been renormalized.
It is now convenient to transform from Eddington-Finklestein 
coordinates back to $(t,r,\theta,\phi)$ coordinates.
Doing so one finds the perturbed metric takes the form:
\begin{equation}
        ds^2 = -[1 + 2\epsilon \rho(r) ] 
		\left(1 - \frac{2m(r)}{r} + \frac{Q^2}{r^2}\right) dt^2
		+ \left(1 - \frac{2m(r)}{r} + 
		\frac{Q^2}{r^2}\right)^{-1} dr^2 + r^2 d\Omega ^2 \; .
\label{pert_rsn}
\end{equation}

The remaining integration constant, $C_2$, was fixed in the case of
the massless field studied in Ref.\cite{AHWY} by enclosing the black
hole in a cavity and imposing microcanonical boundary conditions.
However, as discussed above there is no gas of massive scalar
particles surrounding the black hole, because the DeWitt-Schwinger
approximation is state (and hence temperature) independent.  In the
domain where the DeWitt-Schwinger approximation is valid (i.e., $Mm >
2$), the temperature of the hole is so low that it creates a
negligible number of such particles.  The stress-energy associated
with the massive scalar field in this limit is essentially the result
of vacuum polarization, not particle production.  Hence, placing the
hole in a cavity to allow thermal equilibrium is neither appropriate
nor necessary.

Instead, the integration constant $C_2$ may be fixed by requiring that
$g_{tt}$ in the perturbed metric of Eq.\ (\ref{pert_rsn}) approach the
usual value of $-1$ as $r \rightarrow \infty $. This implies that $C_2$
simply determines the normalization of the time coordinate at infinity.
For $g_{tt}$ to approach $-1$ as a limiting value requires that
$\rho (\infty) = 0$, and therefore that
\begin{equation}
        C_2 = - \tilde{\rho}(r)|_{r \rightarrow \infty} \; ,
\label{C2_def}
\end{equation}
where
\begin{equation}
        \tilde{\rho}(r) = \rho (r) \mid_{C_2 = 0}.
\label{tilde_rho}
\end{equation}
It is worth noting that this condition is identical to the
microcanonical boundary condition in the limit that the cavity
radius approaches infinity. With the fixing of $C_2$ the perturbed
spacetime is now completely defined; $C_1$ and $C_2$ no longer appear
in the perturbed metric, having been fixed; the mass $M$ now refers
to the ``dressed'' black hole mass, defined implicitly by the 
horizon radius through Eq.\ (\ref{r_h_two}).


\section{Results}
\label{sec:results}

In this section we will concentrate on the examination of two properties
of the perturbed black hole metric: first, the relation of the mass
$M$ defined by the horizon radius to the mass that would be measured at
infinity by a Keplerian orbit, $M_{\infty}$; second, the effect of
the semiclassical perturbation on the temperature of the black hole.

With the integration constant $C_1$ absorbed into the horizon-defined
mass $M$ as described in Eq.\ (\ref{newm}) above, the horizon radius
keeps its simple, Reissner-Nordstr\"{o}m form, as seen in Eq.\
(\ref{r_h_two}).  However, the price paid is that now the mass $M$ is
not the mass that would be measured for the perturbed black hole by
an observer at infinity, say by observing the properties of an
orbiting test mass at large $r$.  The mass of the black hole at
infinity will be
\begin{equation}
M_{\infty} = M \left[1+\epsilon \mu(r)|_{r \rightarrow \infty} \right]
= M + \delta M \; .
\label{minfty}
\end{equation}
The difference between the mass measured at infinity and the horizon
defined mass is then
\begin{eqnarray}
\delta M &=& \frac{\epsilon M}{\pi m^2}\left[- \frac{4169 M^4}{158760 r_{+}^6}
	+\frac{461 M^3}{6615 r_{+}^5} - \frac{6607 M^2}{105840 r_{+}^4} + 
	\frac{3007 M}{158760 r_{+}^3}
\right. \nonumber \\ \nonumber \\
& & {} \left.
        +\xi \left(\frac{4 M^4}{45 r_{+}^6} - \frac{11M^3}{45r_{+}^5} +
         \frac{41 M^2}{180 r_{+}^4}
        - \frac{13 M}{180 r_{+}^3} \right) \right] \; .
\label{deltaM}
\end{eqnarray} 

Depending on the value of the scalar field's curvature coupling,
$\xi$, $M_{\infty}$ can be either larger ($\delta M > 0$) or smaller
($\delta M < 0$) than $M$.  Examination of Eq.\ (\ref{deltaM}) shows
that for a conformally coupled field ($\xi = 1/6$), $\delta M > 0$ for
all values of $Q^2/M^2 \lesssim 0.954463$, while for a minimally
coupled field ($\xi = 0$), $\delta M > 0$ for all $Q^2/M^2 \lesssim
0.998701$.  Interestingly, in the extreme Reissner-Nordstr\"{o}m
limit, for which $Q^2/M^2 =1$, $\delta M$ becomes negative and
independent of $\xi$;
\begin{equation}
	\delta M_{ERN} = \frac {-17\epsilon}{317520 \pi m^2 M} \; .
\label{deltaMERN}
\end{equation}
This implies that an extreme black hole, perturbed semiclassically by
a massive quantized scalar field, will have a charge-to-mass ratio (as
measured at infinity) greater than unity.

In order to determine the effect the presence of the quantized
scalar field has on the temperature, the surface gravity of the black
hole must be calculated.  For the 
perturbed metric in Eq.\ ({\ref{pert_rsn}) the surface gravity to first
order in $\epsilon$ is
\begin{equation}
        \kappa = \frac{\sqrt{M^2-Q^2}}{r_+^2}\left( 1 + \epsilon C_2 \right)
        +4 \pi r_+ \langle T_t^t \rangle \; ,
\label{kappa_rsn}
\end{equation}
which reduces to the usual Reissner-Nordstr\"{o}m surface gravity as
$\epsilon \rightarrow 0$.  The perturbation in the surface gravity is
given by
\begin{equation}
        \delta \kappa = \kappa - \frac{\sqrt{M^2-Q^2}}{r_+^2} \; ,
\label{delta_kappa}
\end{equation}
or, explicitly,
\begin{eqnarray}
\delta \kappa & = & \frac{\epsilon M^2}{35280 \pi m^2 r_{+}^8} 
	\left[ 2458 M^3 -5766 M^2 r_{+} + 4617 M r_{+}^2 -1239 r_{+}^3
	\right. \nonumber \\ \nonumber \\ & &  {} \left.
	+ \xi \left( -9408 M^3 + 22736 M^2 r_{+} -18620 M r_{+}^2 
	+ 5096 r_{+}^3 \right) \right] \; .
\label{dkappa}
\end{eqnarray}
The perturbation in the surface gravity depends on the value of the
curvature coupling constant, $\xi$, which appears in both the
expression for $C_2$ and $\mu$.  In addition the surface gravity will
also depend on the field mass $m$ and the overall size of the
perturbation, $\epsilon$.  Both of these are multiplicative factors
that only affect the overall size of the perturbation to the surface
gravity.  Our attention here is focused on:  (1) the sign of $\delta
\kappa$ for various combinations of black hole state and field
curvature coupling, and (2) which black holes states can conceivably
have the total surface gravity ($\kappa$, in the perturbed state), and
hence temperature, equal to zero.

Since the expression for $\delta \kappa$ is linear in $\xi$, it is
a simple matter to find the value of $\xi$ for each value of $Q^2/M^2$
that will result in $\delta \kappa = 0$. The domain of allowed black
hole states may then be divided into regions where $\delta \kappa > 0$,
and regions where $\delta \kappa < 0$. Figure \ref{dkappafig} illustrates
the sign of $\delta \kappa$ as a function of $Q^2/M^2$ and $\xi$.

In the Schwarzschild limit, $\delta \kappa$ simplifies to the form
\begin{equation}
	\delta \kappa_{\rm Sch} = \epsilon \; \left(\frac{-37 + 168 \xi}
	{645120 \pi m^2 M^3} \right) \; ,
\label{dkappaSCH}
\end{equation}
which is negative for both the minimal and conformally coupled
massive scalar field, as well as for any field with $\xi < 37/168$.
Thus, the likely effect of a semiclassical perturbation from a
massive quantized scalar field on a Schwarzschild black hole is
to lower its temperature.

In the extreme Reissner-Nordstr\"{o}m limit, $\delta \kappa$ becomes
\begin{equation}
	\delta \kappa_{\rm ERN} = \epsilon \; \left( \frac{5-14 \xi}
	{2520 \pi m^2 M^3} \right) \; ,
\label{dkappaERN}
\end{equation}
which is positive for both the minimal and conformally coupled field,
as well as for any field with $\xi < 5/14$.  Thus, for a
Reissner-Nordstr\"{o}m black hole with a semiclassical perturbation
provided by a massive quantized scalar field, unless the curvature
coupling takes on apparently unnatural values ($\xi > 5/14$), there
will be {\it no} zero temperature solution.  The extreme black hole
(now defined as the black hole with maximum possible charge-to-mass
ratio, beyond which are naked singularity solutions) will have a
nonzero temperature.

Given the somewhat surprising result that zero temperature solutions
to the linearized semiclassical backreaction equations do not exist
for realistic values of the curvature coupling constant $\xi$, it is
perhaps useful to ask whether zero temperature solutions to the full
nonlinear semiclassical backreaction equations exist.  Although it is
difficult to find solutions to the nonlinear equations everywhere
outside the event horizon (even with the use of the DeWitt-Schwinger
approximation) it is possible to solve the equations near the event
horizon.  To do so one can simply expand the metric functions, the
Einstein tensor, and the stress-energy tensor in powers of $(r-r_h)$
and solve the equations order by order in $(r-r_h)$.  Utilizing
this approach with the full nonlinear equations, we have found that
zero temperature solutions of the extreme Reissner-Nordstr\"{o}m form near
the event horizon exist.  These solutions have a slightly different
ratio between the charge $Q$ and the radius $r_h$ of the event horizon
than do the classical Reissner-Nordstr\"{o}m solutions.

However, the existence of zero temperature local solutions to the full
nonlinear equations is probably irrelevant from a physical point of
view.  The reason is that the DeWitt-Schwinger approximation contains
terms with up to six derivatives of the metric.  These higher
derivatives lead to many locally (i.e., near the horizon) sensible
solutions to the equations that are an artifact
of the approximation since the exact stress-energy tensor contains
terms with up to only four derivatives of the metric. Even in the case 
when the exact stress-energy
tensor is used in the semiclassical backreaction equations it has been 
argued that the higher
derivatives here also lead to physically unacceptable
solutions~\cite{Parker_Simon}.  Thus the most likely situation is that
the only solutions to the nonlinear equations that are physically
acceptable when the DeWitt-Schwinger approximation is used are those
that reduce to the solutions to the linearized equations in the limit
$mM \rightarrow \infty$.  In this case we have already seen that zero
temperature solutions do not exist for reasonable values of $\xi$ .


\section{Discussion}
\label{sec:summary}

We have investigated the effect the vacuum stress-energy of a
quantized massive scalar field has on the geometry of a charged black
hole, within the context of linear perturbation theory.  We have found
the metric functions that describe such a semiclassically perturbed
black hole, to first order in $\epsilon = \hbar/M^2$.  We have shown
that the mass of such a black hole, as measured at infinity, will
differ from the mass defined in terms of the horizon radius.  For an
extreme black hole, the mass at infinity will always be less than the
mass defined by the horizon radius.  The charge-to-mass ratio of an
extreme black hole, as measured at infinity, will exceed unity for all
values of the massive field scalar curvature coupling.  We have also
examined the effect of the semiclassical perturbation on the surface
gravity of the black hole.  For reasonable values of the scalar field
curvature coupling, the perturbation lowers the temperature of a
Schwarzschild black hole.  For an extreme black hole, the temperature
is raised by the perturbation for any scalar field with $\xi < 5/14$,
including the physically interesting cases of minimal and conformal
coupling.  Thus, within the context of Reissner-Nordstr\"{o}m black
holes semiclassically perturbed by the vacuum energy of a massive
scalar field, there are no plausible zero-temperature solutions.

\acknowledgments

This work was supported in part by National Science Foundation Grant
No. PHY-9734834 at Montana State University and No. PHY-9800971 at
Wake Forest University.


\newpage

\begin{figure}
	\caption{The curves represents semiclassical black hole 
        solutions for which the change in temperature is zero for
        particular values of the charge and curvature coupling
        constant.}
\label{dkappafig}
\end{figure}

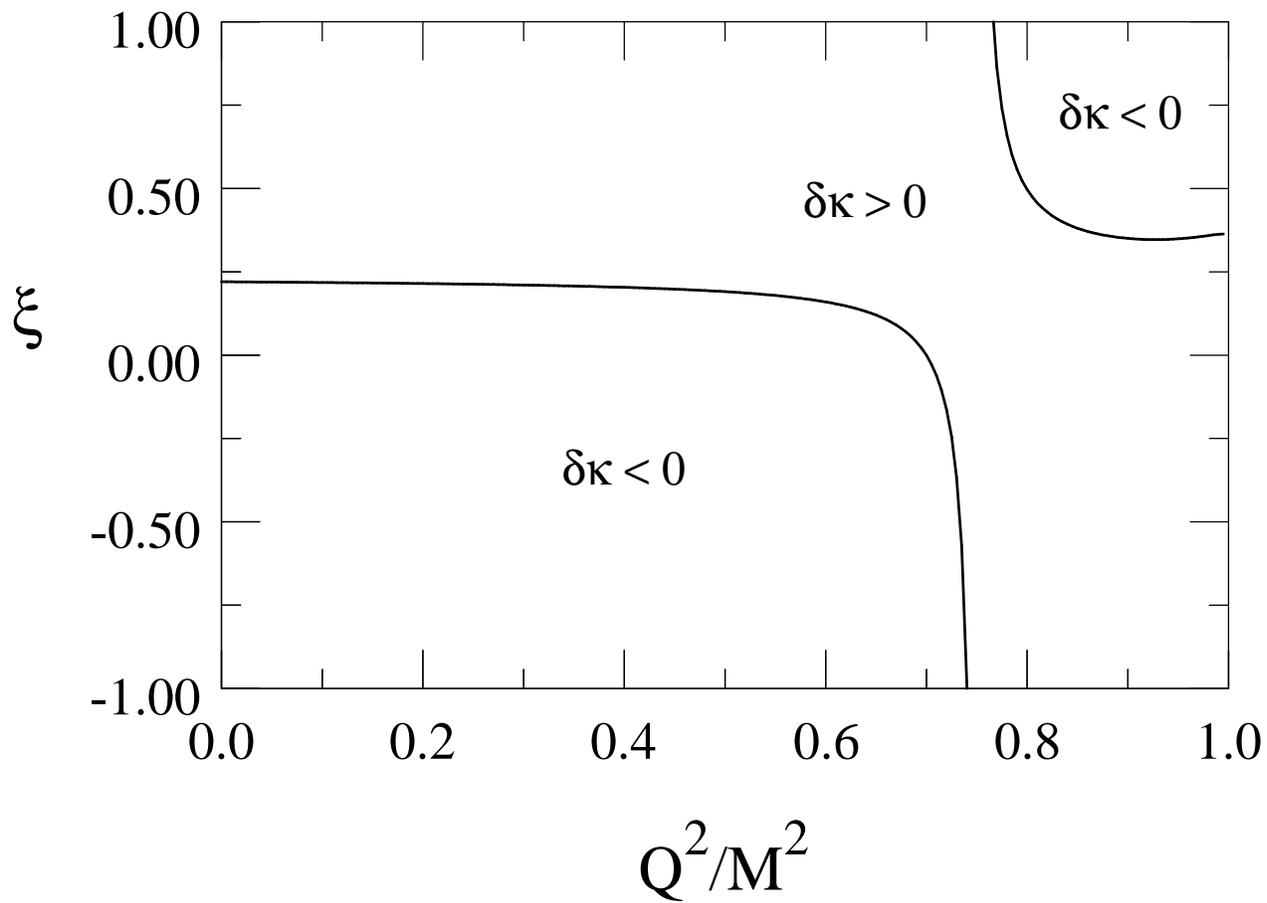
\end{document}